\documentclass{emulateapj}

\usepackage[utf8]{inputenc}
\usepackage{amsmath}
\usepackage{graphicx}
\usepackage{amsmath} 
\usepackage{amssymb}  
\usepackage{amsthm}
\usepackage{color}
\usepackage{url}

\definecolor{dkgreen}{rgb}{0,0.6,0}
\definecolor{gray}{rgb}{0.5,0.5,0.5}
\definecolor{mauve}{rgb}{0.58,0,0.82}

\newcommand{\tamas}[1]{\textcolor{dkgreen}{#1 [Tamas]}}

\newcommand{\like}{\mathcal{L}}  
\newcommand{\mlike}{\mathcal{M}}
\newcommand{\partn}{\mathcal{P}}

\begin{document}





\title{Probabilistic cross-identification of multiple catalogs in crowded fields}

\iftrue

\author{
Xiaochen Shi\altaffilmark{1,4},
Tam\'{a}s Budav\'{a}ri\altaffilmark{1,2,3},
\and
Amitabh Basu\altaffilmark{1}
}
\altaffiltext{1}{Dept.~of Applied Mathematics and Statistics, The Johns Hopkins University}
\altaffiltext{2}{Dept.~of Computer Science, The Johns Hopkins University}
\altaffiltext{3}{Dept.~of Physics and Astronomy, The Johns Hopkins University}
\altaffiltext{4}{Dept.~of H. Milton Stewart School of Industrial and Systems Engineering, Georgia Institute of Technology}

\shortauthors{Shi, Budav\'ari \& Basu}

\shorttitle{Cross-Identification of Multiple Catalogs in Crowded Fields}

\else

\author[ams]{Xiaochen Shi}
\author[ams,cs,pha]{Tam\'{a}s Budav\'{a}ri}
\author[ams]{Amitabh Basu}

\address[ams]{Dept.~of Applied Mathematics \& Statistics, Johns Hopkins University, USA}
\address[cs]{Dept.~of Computer Science, Johns Hopkins University, USA}
\address[pha]{Dept.~of Physics \& Astronomy, Johns Hopkins University, USA}

\fi

\begin{abstract}
Matching astronomical catalogs in crowded regions of the sky is challenging both statistically and computationally due to the many possible alternative associations. \citet{BB16-hungarian} modeled the two-catalog situation as an {\em Assignment Problem} and used the famous Hungarian algorithm to solve it. Here we treat cross-identification of multiple catalogs by introducing a different approach based on integer linear programming. We first test this new method on problems with two catalogs and compare with the previous results. We then test the efficacy of the new approach on problems with three catalogs. The performance and scalability of the new approach is discussed in the context of large surveys.
\end{abstract}



\section{Motivation}  

Dedicated telescopes systematically survey the night sky often observing the same area. 
These exposures together enable multi-wavelength and time-domain studies. 
A central problem with multiple observations is to reconcile the different data sets that correspond to the same part of the sky and build a consensus for the parameters of the observations. 
This problem is generally known as {\em cross-identification} or {\em catalog matching}. 
Recently, significant progress has been made in the development of statistical and computational methods. 
Based on Bayesian hypothesis testing, \citet{BS08-BayesCrossID} developed a framework which proved to be applicable in a variety of scenarios, e.g., to account for proper motion \citep{kerekes}, for radio morpholgy \citep{dongwei2015radio} or galaxy clustering \citep{mallinar-ascom201783}; also see the review by \citet{budavari_loredo} and references within. 
New methods were also created to accelerate and automate the matching process on the largest catalogs available using spatial indices, e.g., the Hierarchical Triangular Mesh \citep[HTM;][]{htm} or HEALPix \citep{healpix} within databases 
\citep[e.g., the SkyQuery,][]{budavari_skyquery:_2013} and even on Graphics Processing Units \citep{matthias2017gpu}.
While these methods have opened the door to statistically sound data fusion techniques, they considered associations in isolation and simply assessed their quality without a global context.

In \citet{BB16-hungarian}, a non-trivial step beyond previous approaches was taken by introducing methods that maximized the likelihood of a global catalog matching, as opposed to a greedy choice of local likelihood maximization of isolated tuples. This was achieved by borrowing tools from combinatorial optimization which provide efficient algorithms for solving this much more difficult global optimization problem.
Even so, the approach of \citet{BB16-hungarian} can only be applied to matching two catalogs at a time, which is a serious limitation, as matching more catalogs is often needed in practice. 
In these situations, only heuristic methods have been available that find solutions based on greedy, locally good associations, and may not provide globally optimal matchings. This serious defect results in poor performance especially when matching a large number of catalogs in crowded fields. 

In this paper, we introduce a new class of algorithms that efficiently solves the problem of associating multiple catalogs while guaranteeing global optimality with respect to precise statistical objectives. 
Section~\ref{sec:cat-match} discusses the underlying partition models, the likelihood functions, marginal likelihoods and Bayes factors, as well as our new optimization procedure. 
In Section~\ref{sec:test}, we study realistic simulations to test the accuracy and applicability of the new method. We also compare with the results of \citet{BB16-hungarian} using the Hungarian algorithm \citep{Munkres57hunalg}. 
In Section~\ref{sec:disc}, we demonstrate the improvements achieved by the new approach on three catalogs using a suite of simulations.


\section{Catalog-Level Matching}\label{sec:cat-match}

Probabilistic catalog-level cross-identification aims at finding the globally optimal catalog matches in which all associations are valid and no detection appears in multiple matches. Following \citet{budavari_loredo}, we discuss the formal model for the association problem, and introduce the statistical objective we shall use to find the best solution.

\subsection{The Partition Model}
\label{sec:parmo}
We consider the union of all observations across the different catalogs and aim to partition this set of all observations into disjoint subsets.
A subset with multiple elements should be interpreted to represent the hypothesis that all these elements in the subset are observations of the same object in the sky; such a subset is called an {\em association}. Thus, a subset containing a single element is to be thought of as the only observation amongst all the catalogs for that object; such a singleton association is called an {\em orphan}. Clearly, the partitioning cannot be arbitrary: any subset in the partition should contain at most one observation from each catalog (representing the fact that all these elements are observations of the same object in the sky). The fact that we have a partition into {\em disjoint} subsets encodes the fact that an observation should not stand for two different objects in the sky.

\begin{figure}
\centering
\includegraphics[width = 0.45\textwidth, height=0.14\textwidth, trim=13mm 0 11mm 0, clip]{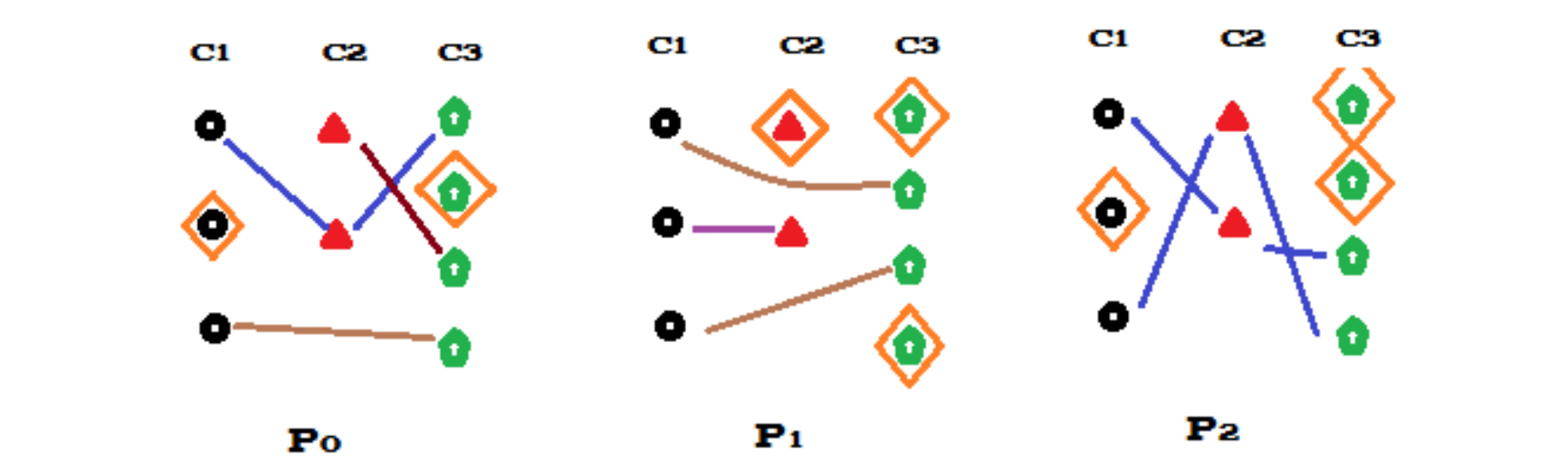} 
\caption{Example association graphs for three partitions $\partn_{0}$, $\partn_{1}$ and $\partn_{2}$ 
in \mbox{eq.(\ref{Partition})}. The nodes shown in different colors (and glyphs) represent sources in three catalogs. The lines correspond to possible associations, and the orange diamonds around the glyphs indicate standalone ``orphan'' detections.}
\label{fig:partition}
\end{figure}

Let us consider a simple case. Suppose there are three catalogs, and they contain three sources, two sources and four sources respectively. The $\left(i,c\right)$ pair denotes the $i$th source in catalog $c$. Let $D$ denote the set of all observations $\left(i,c\right)$, that is \mbox{$D=\lbrace\left(i,c\right) : \forall i,c \rbrace$}. Then a matching between the three catalogs corresponds to a partition of $D$ into disjoint subjects satisfying the constraint outlined in the previous paragraph. For example,
%
%
\iffalse
\begin{equation} \label{Partition}
\begin{aligned}
&\partn_{0}=\Big\lbrace
\big\lbrace\left(1,1\right),\left(2,2\right),\left(1,3\right)\big\rbrace,
\big\lbrace\left(3,1\right),\left(4,3\right)\big\rbrace,\\ 
&\ \ \ \ \ \ \ \ \ \ \ 
\big\lbrace\left(1,2\right),\left(3,3\right)\big\rbrace,
\big\lbrace\left(2,1\right)\big\rbrace,
\big\lbrace\left(2,3\right)\big\rbrace
\Big\rbrace\\
&\partn_{1}=\Big\lbrace
\big\lbrace\left(1,1\right),\left(2,3\right)\big\rbrace,
\big\lbrace\left(3,1\right),\left(3,3\right)\big\rbrace,
\big\lbrace\left(2,1\right),\left(2,2\right)\big\rbrace,\\ 
&\ \ \ \ \ \ \ \ \ \ \ 
\big\lbrace\left(1,2\right)\big\rbrace,
\big\lbrace\left(1,3\right)\big\rbrace,
\big\lbrace\left(4,3\right)\big\rbrace
\Big\rbrace\\
&\partn_{2}=\Big\lbrace
\big\lbrace\left(1,1\right),\left(2,2\right),\left(3,3\right)\big\rbrace,
\big\lbrace\left(3,1\right),\left(1,2\right),\left(4,3\right)\big\rbrace,\\
&\ \ \ \ \ \ \ \ \ \ \
\big\lbrace\left(2,1\right),\left(1,3\right)\big\rbrace,
\big\lbrace\left(4,3\right)\big\rbrace
\Big\rbrace\\
\end{aligned}
\end{equation}
\else
\begin{eqnarray} \label{Partition}
\partn_{0} &=& \Big\lbrace
\big\lbrace\left(1,\!1\right),\left(2,\!2\right),\left(1,\!3\right)\big\rbrace,
\big\lbrace\left(3,\!1\right),\left(4,\!3\right)\big\rbrace, \\ 
& & \ \ 
\big\lbrace\left(1,\!2\right),\left(3,\!3\right)\big\rbrace,
\big\lbrace\left(2,\!1\right)\big\rbrace,
\big\lbrace\left(2,\!3\right)\big\rbrace
\Big\rbrace 
\nonumber \\
\partn_{1} &=& \Big\lbrace
\big\lbrace\left(1,\!1\right),\left(2,\!3\right)\big\rbrace,
\big\lbrace\left(3,\!1\right),\left(3,\!3\right)\big\rbrace,
\big\lbrace\left(2,\!1\right),\left(2,\!2\right)\big\rbrace,
\nonumber \\
& & \ \ 
\big\lbrace\left(1,\!2\right)\big\rbrace,
\big\lbrace\left(1,\!3\right)\big\rbrace,
\big\lbrace\left(4,\!3\right)\big\rbrace
\Big\rbrace
\nonumber \\
\partn_{2} &=& \Big\lbrace
\big\lbrace\left(1,\!1\right),\left(2,\!2\right),\left(3,\!3\right)\big\rbrace,
\big\lbrace\left(3,\!1\right),\left(1,\!2\right),\left(4,\!3\right)\big\rbrace,
\nonumber \\
& & \ \ 
\big\lbrace\left(2,\!1\right)\big\rbrace,\big\lbrace\left(1,\!3\right)\big\rbrace,
\big\lbrace\left(2,\!3\right)\big\rbrace
\Big\rbrace
\nonumber 
\end{eqnarray}
\fi
where $\partn_{0}$ gives five associations representing five objects (with two orphans),  $\partn_{1}$ gives six associations for six objects (with three orphans), and $\partn_{2}$ represents five separate objects (with three orphans). As mentioned earlier, the partition is valid if and only if for any association in the partition has at most one of each catalog. Formally, if it contains a pair \mbox{$\left\{(i_{1},c_{1}),\,(i_{2},c_{2})\right\}$}, then $c_{1}\neq c_{2}$. In this example, there are a total of 7 valid partitions.  We will use \mbox{$o\in{}O_{\partn}$} to denote collection of indices that index the subsets in a valid partition $\partn$, corresponding to objects in $\partn$. The catalog cross-identification problem is then equivalent to finding the optimal partition that has the maximum likelihood.

\subsection{Probabilistic Formalism}


The probabilistic formalism for cross-identification based on Bayesian hypothesis testing lends itself to hierarchical modeling and catalog-level generalizations, however, until now, no approach has been available to efficiently solve the problem of partitioning the sources into globally optimal associations of objects.

Given a valid partition $\partn$, one considers the conditional probability density of the data $D=\lbrace\left(i,c\right)\rbrace$ given $\partn$, i.e., $p\left(D \mid \partn\right)$. The hierarchical Bayesian framework has a vital feature that, given a certain partition, the data from distinct associations are conditionally independent. Therefore, the likelihood function at a certain partition $\partn$ can be factored as
\begin{equation}\label{L(P)}
\like(\partn) \equiv p\left(D \mid {\partn}\right) = \prod_{o\in{}O_{\partn}}\!\!\mlike_{o}
\end{equation}
where $\mlike_{o}$ is the marginal likelihood of object $o\in O_{P}$ in the partition $\partn$ \citep{budavari_loredo}, given by 
\begin{equation}\label{eq:Mo}
\mlike_{o}=\int d\omega \ \rho_{C\left(o\right)}\!\!\left(\omega\right) \prod_{\left(i,c\right)\in S_{o}}\!\!\ell_{ic} \left(\omega \right)
\end{equation}
where $S_{o}$ denotes the subset $\lbrace\left(i,c\right)\rbrace$ in the partition $\partn$ associated with object $o$,  $C\left(o\right)$ represents a collection of catalogs containing sources associated with object o, $\ell_{ic}\left(\omega \right)$ is the likelihood function of source $i$ in catalog $c$ for direction $\omega$, and $\rho_{C\left(o\right)}\left(\omega\right)$ is the prior probability density for  $C\left(o\right)$ depending on the angular and radial selection functions of catalogs.
Our goal is to find the optimal partition by maximizing the $\like(\partn)$ catalog-level likelihood.

The marginal likelihood for non-association given all observed sources in $S_{o}$ is
\begin{equation}
\mlike_{o}^{\textrm{NA}}= \prod_{\left(i,c\right)\in S_{o}} \int d\omega\ \rho_{c}\!\left(\omega\right)\,\ell_{ic} \left(\omega\right)
\end{equation} 
where $\rho_{c}\!\left(\omega\right)$ is the  prior probability density of direction $\omega$ for  an ``orphan'' in catalog $c$. 
%
The Bayes factor is the ratio of the marginal likelihoods for the association and non-association hypothesis, quantifying how much the data supports a specific association $o$, 
\begin{equation}\label{BayF}
B_{o}\equiv \dfrac{\mlike_{o}}{\mlike_{o}^{\textrm{NA}}}\,.
\end{equation}
Large $B_{o}$ denotes that the sources in $S_{o}$ are more likely from same object $o$ than different objects, while $B_{o}$ less than one indicates the sources are more likely from separate objects. 

Considering 
that the product of the marginal likelihoods $\prod \mlike_{o}^{\textrm{NA}}$ over all $o\in{}O_{\partn}$ consists of the same terms 
corresponding to every observed source $(i,c)$,  
\begin{equation}\label{Mnacons}
\prod_{o\in{}O_{\partn}} \mlike_{o}^{\textrm{NA}} 
= \prod_{(i,c)\in{}D}  \int\!d\omega \ \rho_{c}(\omega)\ \ell_{ic}(\omega) 
\end{equation}
%
its value is actually independent of the $\partn$ partition. This means that maximizing $\prod B_{o}$ is equivalent to maximizing the product of the marginals, $\prod\mlike_{o}$.

In practice, we have to choose a member likelihood function, which is often assumed to be Gaussian.
A more general choice is the \citet{fisher} distribution, which is a spherical analog to the normal distribution, 
\begin{equation} \label{eq:fisher}
f\left(x;\omega,\kappa\right)=\dfrac{\kappa}{4 \pi\,\sinh \kappa} \ \exp\big(\kappa\, \omega x\big)
\end{equation}
where $\omega$ is the unit vector of the mode, and $\kappa$ is the concentration parameter. 
We can apply this spherical normal distribution to analytically calculate the Bayes factors using $\ell_{ic} \left(\omega \right) = f(x_{ic}; \omega, \kappa_{ic})$ with observed directions $x_{ic}$. For large concentrations ($\kappa\!\gg\!1$), the flat-sky approximation corresponds to using Gaussian likelihoods with \mbox{$\kappa=1/\sigma^2$}, and the Bayes factor becomes 
\begin{equation} \label{Bay}
B_o = 2^{|S_o|-1} \dfrac{\prod_{ic} \kappa_{ic}}{\sum_{ic}\kappa_{ic}}\!\exp\!\left(-\dfrac{\sum_{ic} \sum_{jk} \kappa_{ic} \kappa_{jk}\psi_{ic,jk}^{2}}{4\sum_{ic} \kappa_{ic}}\right)
\end{equation}
%
%
where the $\prod_{ic}$ and both $\sum_{ic}$ and $\sum_{jk}$ go over $S_o$ set, e.g., \mbox{$(i,c)\in{}S_o$}, the $\kappa_{ic}$ is the concentration parameter that corresponds to the astrometric uncertainty as in eq.(\ref{eq:fisher}), and $\psi_{ic,jk}$ is the angle between the directions of sources $(i,c)$ and $(j,k)$; see \citet{BS08-BayesCrossID}.

In preparation for the optimization, we first take the logarithm of the above objective function $\prod B_{o}$, which yields a minimization problem of the sum
\begin{equation}\label{minB}
F(\partn) = -\sum_{o\in O_{\partn}} \ln B_{o}
\end{equation}
over all possible partitions $\partn$ that satisfy the condition stated earlier that if $(i_1,c_1)$ and $(i_2,c_2)$ belong to the same subset in the partition, then $c_1 \neq c_2$.



\begin{figure}
\epsscale{1.15}
\plotone{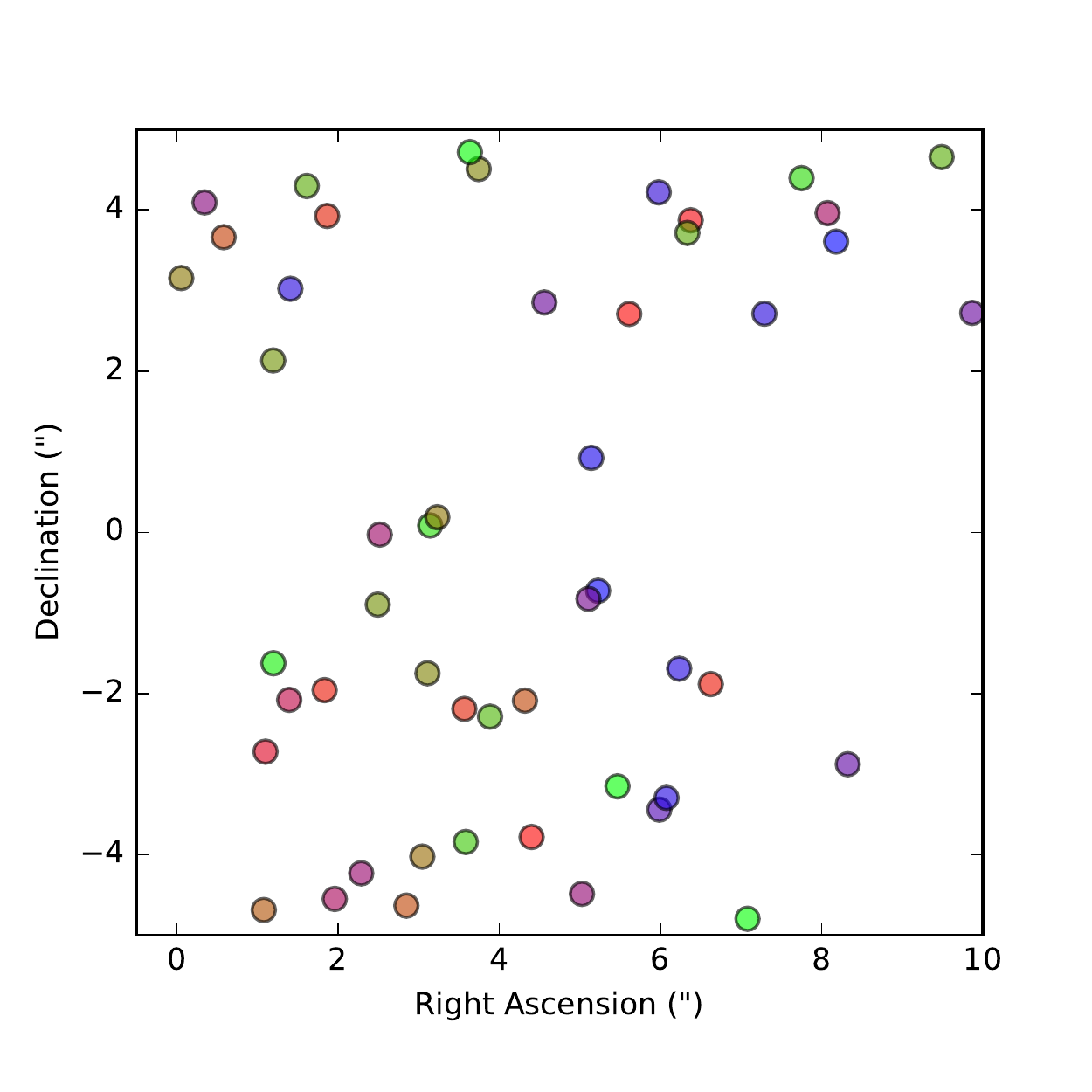} 
\caption{The mock universe consists of objects with the randomly chosen directions and intrinsic properties shown by colors. This dense field is generated for visualization purposes.}
\label{fig:mock}
\end{figure}

\begin{figure*}
\centering
\includegraphics[width = 0.99\textwidth,trim=15mm 25mm 15mm 25mm, clip]{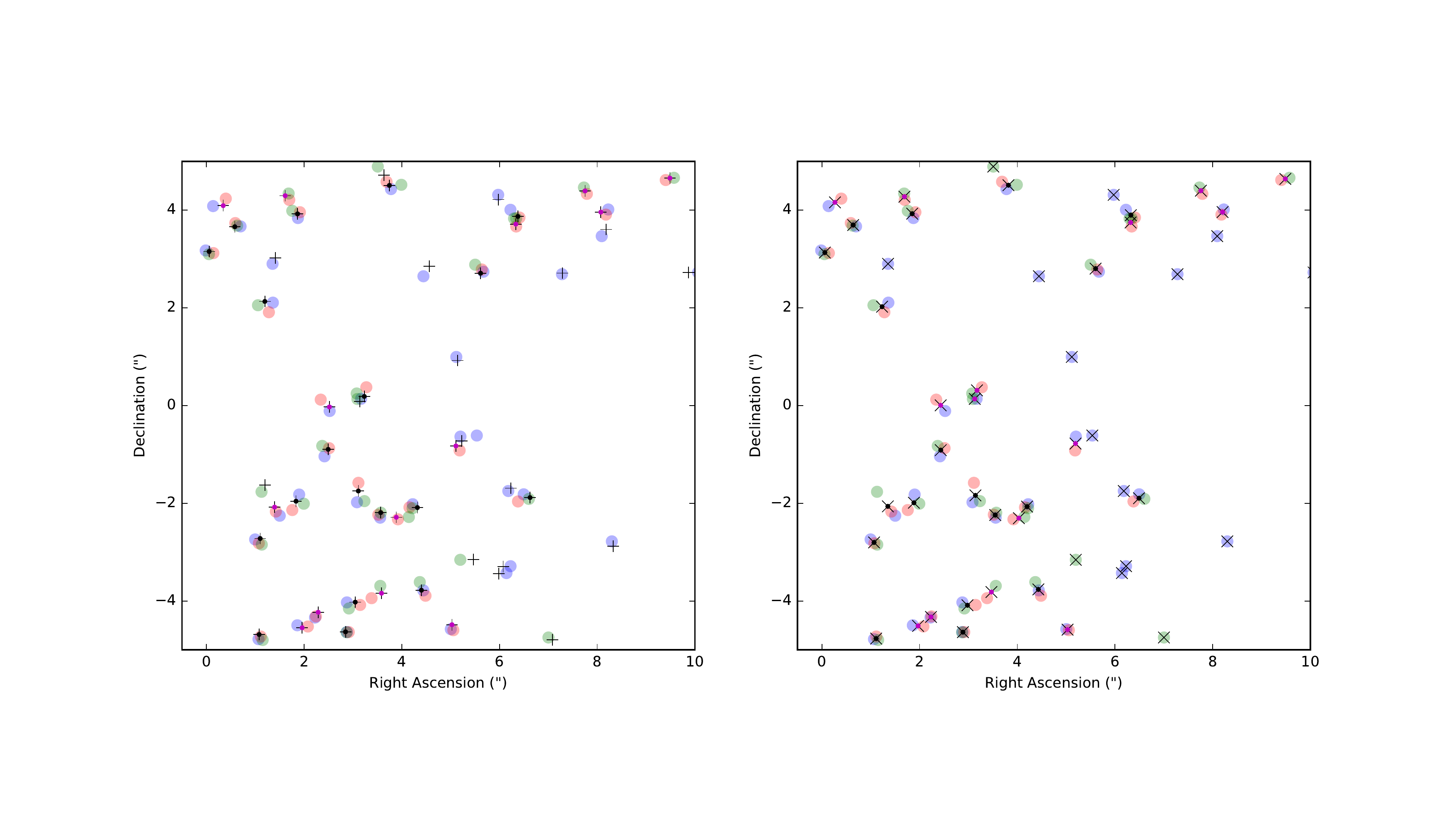}   
\caption{
Simulated sources to illustrate the difficulties in crowded fields.
\textit{Left:} The ground truth. Large green, blue and red dots  are sources from  three catalogs and the ``$+$'' signs mark the true directions of the objects. The small solid dots in magenta represent objects observed by two catalogs and the black dots are  those captured by  all three catalogs.
\textit{Right:} matching result. The ``$\times$'' marks show the estimated objects. The small black solid dots are 3 catalog associations, while small magenta dots  represent estimated associations between every two catalogs.}
\label{fig:truth}
\end{figure*}

\subsection{Solution by Integer Linear Programming}\label{sec:ILP}

We will solve the problem of minimizing~\eqref{minB} by formulating it as an {\em integer linear program} or ILP for short. An ILP is a mathematical optimization model that consists of several decision variables whose values are constrained to take only integer values, and the variables are subject to linear equality or inequality constraints. Subject to these integrality and linear constraints, the goal is to find values for the variables that minimizes a given linear function of these decision variables.

To model~\eqref{minB} as an ILP, we introduce variables for each possible valid subset that can be formed from the given data set $D$ as described in Section~\ref{sec:parmo}. For every nonempty subset \mbox{$T \subseteq D$} that contains at most one element from each catalog, we introduce a variable $x_T$ that will be constrained to be integer, and also by the linear inequalities \mbox{$0 \leq x_T \leq 1$}. This forces $x_T$ to be a {\em binary variable} that takes only values $0$ or $1$. 
The interpretation of this is the following: if \mbox{$x_T = 1$} in a solution, then the subset $T$ is taken as a subset in the partition, and if \mbox{$x_T = 0$}, then the subset $T$ does not appear in the partition. 
To ensure that we indeed have a partition, we must impose the linear equality constraint $\sum x_T = 1$ for every element \mbox{$(i,c)\in{}D$} where the summations runs over all $T$ that contains $(i,c)$.
This forces exactly one subset to contain the element $(i,c)$, and thus the collection of subsets $T$ such that $x_T=1$ will form a valid partition. Note that we have already modeled the fact that if $(i_1, c_1)$ and $(i_2,c_2)$ are in the same subset then $c_1 \neq c_2$, by introducing decision variables only for those subsets of $D$ that contain at most one element from each catalog. 

To summarize, any valid partition $\mathcal{P}$ corresponds to an assignment of values to the variables $x_T$ that satisfy the following constraints:
%
%
\begin{eqnarray}\label{eq:constraints}
x_T \in \mathbb{Z}\,, & 0 \leq x_T \leq 1\,, \\
\sum_{T:(i,c)\in{}T} \!\!\! x_T = 1 & \quad \textrm{for every }(i,c) \in D\,.  \nonumber
\end{eqnarray}
Conversely, any $\bar{x}$ satisfying all of these constraints provides a valid partition as the collection of subsets $\{T: \bar{x}_T=1\}$.

The next piece is to model the objective function 
$F(\partn)$
using our decision variables $x_T$.
This is a simple task achieved by computing the constants 
\mbox{$w_T = -\ln {B}_{T}$} for every subset $T$ (we abuse notation slightly by using $B_T$ to denote $B_o$ where $o$ is the ``potential" object represented by the subset $T$). Then the objective we want to minimize is simply $\sum w_T x_T$.

As a final simplification step, we observe that the number of variables can be significantly reduced for efficiency. The key point is that \mbox{$\ln {B}_{o}=0$} if the object $o$ is an orphan, which follows from the definition of the Bayes factor.
Thus, these coefficients do not contribute to the objective function. 
So one can simply remove the variables $x_T$ such that \mbox{$|T|=1$}, i.e., $T$ is a singleton
, and replace the partition constraint \mbox{$\sum{}x_T\!=\!1$} in eq.\eqref{eq:constraints} by \mbox{$\sum{}x_t\!\leq\!1$}, where again the summation runs over all $T$ that contains $(i,c)$.
%
In the final solution, if an element $(i,c)$ is not part of any subset $T$ such that $x_T=1$, then $(i,c)$ should be considered an orphan. 
Thus, our final integer linear programming function for minimizing eq.\eqref{minB} over all valid partitions $\mathcal{P}$ is 
\iftrue
\begin{eqnarray}\label{Nway}
\min            & \displaystyle \sum_{T} w_{T}\,x_{T} &  \\
\textrm{subject to }  & x_T \in \mathbb{Z} \ \ \textrm{and} \ & 0 \leq x_T \leq 1\quad\textrm{for all }T,  \nonumber \\
\textrm{and} & \displaystyle \sum_{T \ni (i,c)} \!\!\!x_T \leq 1 & \textrm{ for every }(i,c) \in D\,. \nonumber
\end{eqnarray}
\else
\begin{equation}\label{Nway}
\begin{array}{rcl}
\min & \sum_{T} w_{T}\,x_{T}& \\
\textrm{subject to} & \sum_{T : (i,c) \in T} x_T 
\leq 1 & \textrm{for every }(i,c) \in D\\ 
&0 \leq x_T \leq 1 & \\
&x_T \in \mathbb{Z}&
\end{array}
\end{equation}
\fi

In practice, the number of valid partitions can be large but the integer programming algorithms can be further improved with simple thresholding. We can reduce the number of variables by making the observation that the objective coefficient $w_T$ for some subsets $T$ is so large that the corresponding variable cannot possibly be set to $1$ in any optimal solution (since we are minimizing). This typically happens for a subset $\hat{T}$ that  contains sources which are very far away, thus producing a large $w_{\hat{T}}$ value. So, by setting heuristic threshold to rule out subsets with large $w_{T}$, we eliminate a large fraction of the variables and accelerate the integer programming algorithm. This {\em pruning} is critical in practical implementations, and even a threshold of \mbox{$\ln B_o\!=\!0$} can do the trick, which is the value for orphans; thus, if \mbox{$w_{\hat{T}}\!\geq\!0$} for some $\hat{}T$, one can create an equally good or better solution by setting all the sources in $\hat{T}$ as orphans, see also \citet{budavari_basu_1538-3881-152-4-86} for more details.

Many algorithms and off-the-shelf software solutions exist for ILP problems, which are efficient for large-scale problems. In this paper, we employ the commercial \citet{gurobi} package to test the efficacy of our approach. Our implementation is entirely in Python and interfaces with Gurobi via the \texttt{gurobipy} module to create the problem definition and to run the solver.

\section{Mock Objects and Catalogs}\label{sec:test}

We apply the ILP cross-identification procedure to realistic simulations of point sources to study the performance. 
%
We generate mock objects in a small field of view with a given density. Each object is assigned a random direction $\omega$ and a ``physical property'' represented by number $u$ chosen uniformly at random in $\left[0,1\right]$; see also in \citet{heinis}. 
Figure~\ref{fig:mock} shows these mocks with different colors represent the intrinsic properties. 

Next we generate catalogs with directional errors considering astronomical uncertainties. For the illustrations, we take Gaussian errors with a nominal 0.1" standard deviation.  
Because different instruments have different sensitivities, each catalog observes only a subset of objects. This is modeled by assigning a preset interval to each catalog, and only those objects with $u$ values in this interval appear in this catalog. The associations occur when catalogs have overlapping intervals, a common range of selected objects.
For example, the left panel of Figure~\ref{fig:truth} illustrates generated source catalogs with different selections functions. The directions of the detections are plotting in red, green and blue, which scatter around the true direction shown by the ``$+$'' sign; see the detailed analysis later.

First, however, we apply the new ILP algorithm to a suite of 2-way matching problems and test it against the Hungarian Algorithm proposed in \citet{budavari_basu_1538-3881-152-4-86}.
As expected, the outcomes are exactly the same, since the two algorithms are optimizing the same objective. The statistical properties of these associations are discussed in detail by \citet{budavari_basu_1538-3881-152-4-86}.

\subsection{Matching Three Catalogs}
\label{app3}
Matching three catalogs is much more challenging computationally and procedures such as the Hungarian Algorithm are not available. Our new approach, however, works just the same because no assumption on the number of catalogs is needed for the model set up in~\eqref{Nway}.

We implement our approach in Python within a Jupyter Notebook. 
As mentioned above, we use the \citet{gurobi} 
package
to solve the integer linear programs that arise from~\eqref{Nway} for the 3-catalog scenario.


In this application test, both objects density and selection function are accessible from simulation process, making prior probability of each 2-way matching and 3-way matching a known parameter; see details in eq.(25) by \citet{BS08-BayesCrossID}. Thus, by picking a conservative posterior probability threshold, we can easily compute the thresholds for Bayes Factor, producing more accurate estimation. However, the situation is much harder for real data where neither priors nor posteriors are obtained. \citet{budavari_loredo} discuss estimating priors and posteriors in this complicated scenario, which will be a direction for future study.

The right panel of Figure~\ref{fig:truth} illustrates our results. The left panel shows the objects with ``$+$'' signs indicating the true directions on the sky, as well as the noisy (simulated) measurements in the three catalogs shown in red, green and blue. The solid  dots represent objects seen in multiple exposures. In the right panel, the ``$\times$'' signs denote the directions of the estimated objects. The small solid black dots illustrate the positions of objects estimated as associations involving three sources, while the small solid dots in magenta correspond to 2-way associations between every two catalogs. The remaining are orphan associations.

\begin{figure}
\includegraphics[width = 0.95\textwidth,trim=15mm 0mm 5mm 0,clip]{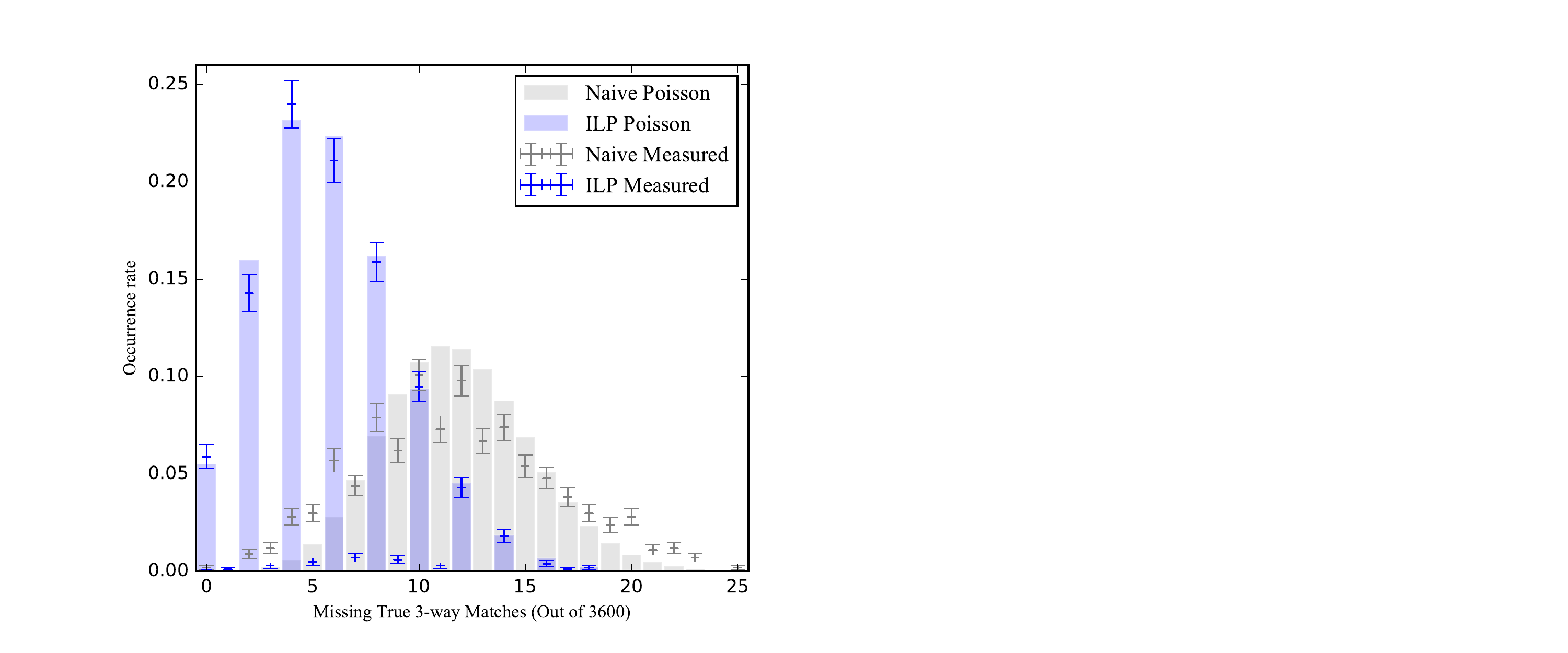}  
\caption{Case 1: three catalogs with identical selection intervals. The blue error bar show the occurrence rate when conducting new algorithm while the gray error bar indicates the rate of naive algorithm. The gray solid bars show Poisson with $\lambda=11.83$ and blue solid bars is Poisson with $\lambda =2.89$. 
}
\label{fig:testC}
\end{figure}

\subsection{Matching Multiple Catalogs}

So far we demonstrated the ILP method on 2 and 3 catalogs, but the new approach is directly applicable to matching \mbox{$N\!>\!3$} catalogs, see eq.\eqref{Nway}.
This, however, does not  mean that the $N$-way problem easy. In principle, the number of variables in the optimization increases dramatically as the number of catalogs grows, which will impact the performance of the ILP solver despite additional possibilities for preprocessing and pruning. This is because the general algorithms for solving ILPs may not be able to take full advantage of the special structure of the problem. Further development and polyhedral study of the system in eq.\eqref{Nway}  could lead to significant speedups over using off-the-shelf software, such as Gurobi.


\begin{figure*}
\centering
\hspace*{-1cm}\includegraphics[width = 1.1\textwidth]{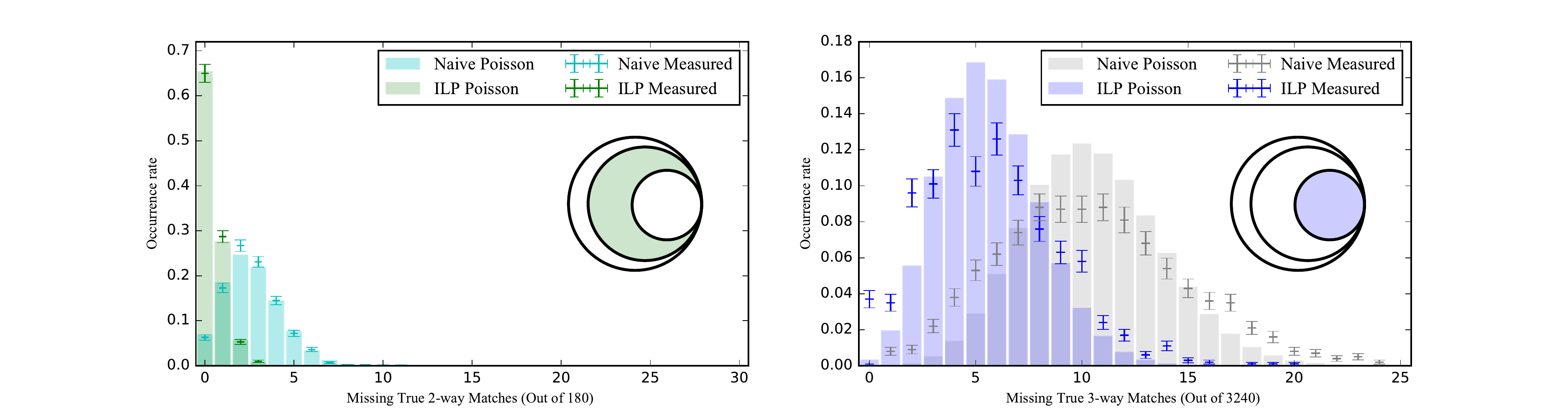}   
\caption{Case 2: three catalogs with varying depths. \textit{Left:} Rate of missing true 2-way matches. The green error bars represent the rate from the new algorithm, which we overplot the green histogram showing the Poisson distribution with $\lambda$ set to the sample mean. The cyan error bars and the histogram corresponds to results obtained by the best possible naive (nearest neighbor) method. \textit{Right:} Rate of missing true 3-way matches. The blue error bars show the result from the new approach, and blue histogram illustrates a Poisson distribution for comparison. The gray error bars and the histogram correspond to results from the naive algorithm.}
\label{fig:testD}
\end{figure*}

\begin{figure*}
\centering
\hspace*{-1cm}\includegraphics[width = 1.1\textwidth]{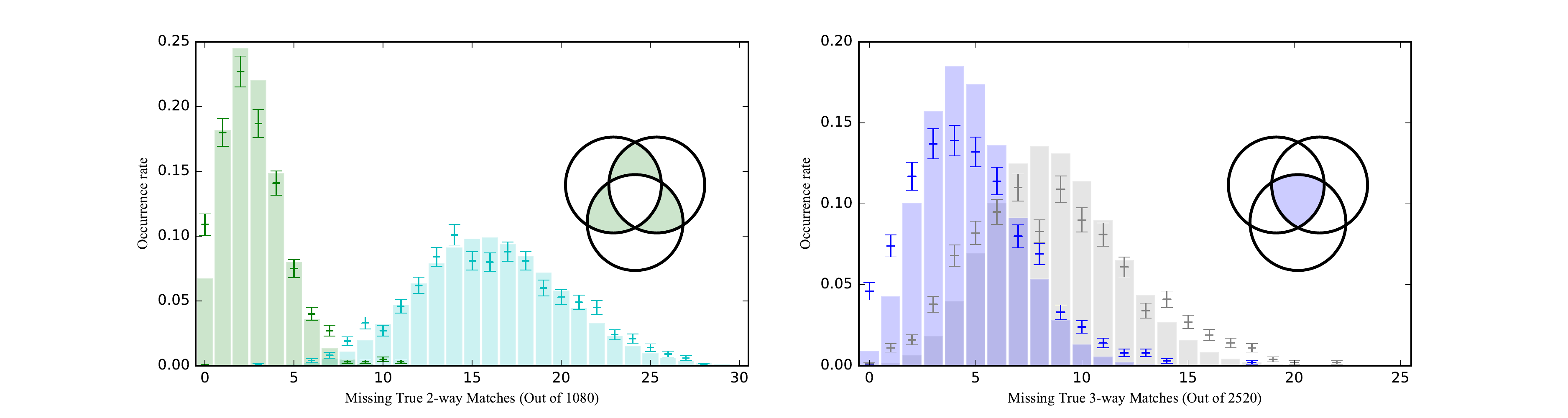}   
\caption{Case 3: three catalogs with different populations. In this example, the virtual surveys have different selection functions, hence, observe different subsets of the (mock) objects. The panels and their content are similar to those in Figure~\ref{fig:testD}.}
\label{fig:testS}
\end{figure*}

\section{Improved Accuracy}
\label{sec:accuracy}

To quantify the new method's statistical accuracy, a series of tests are performed with our ILP algorithm and, as a comparison, a simpler greedy technique.
\color{black}
This naive approach considers all candidate associations with sources closer than a predefined angular separation threshold, and picks the best one based on the average (or maximum) pairwise distances. The match is added to the output catalog and its sources are removed from the catalogs. One can repeat the above greedy step until no candidate associations are left.
%
%
To make the naive method most competitive, we tested multiple thresholds from 4$\sigma$ to 7$\sigma$, where $\sigma$ is the uncertainty of the source directions, and picked the best case performing closest to the ground truth. 

For each mock universe, objects are randomly generated in a 3'$\times$3' field of view with density of 400 objects per square arc minutes.
we took $\sigma$=0.04" as astronomic uncertainty to simulate three catalogs, which were matched by both the ILP and naive methods repeatedly. 
These numbers are somewhat arbitrary but motivated by the expected performance of upcoming surveys, such as the Large Synoptic Survey Telescope (LSST).
\color{black}
In our tests the best naive threshold was $6\sigma$, which is what we adopted for the greedy procedure. 
Next we summarize the findings in three relevant scenarios that differ effectively by the selection functions of the different catalogs.

\subsection{Identical Selection Functions}
\label{identical}
We first consider the idealized scenario of identical observations (and selection intervals), which yields the same sets of objects in the three generated catalogs. If we generated $3600$ mock objects, there should be $3600$ 3-way association formed from the 3 catalogs. 
Figure \ref{fig:testC} presents the distribution of missing matches from both methods. 
The gray error bars represent the measurement of naive method and their square root, which is close to a Poisson distribution with sample mean $\lambda=11.82$ shown by the solid gray bars. 
\textcolor{black}{We see that this method almost never provides perfect results and it typically makes 11-12 mistakes in this small field of view. The results from the ILP method shown in blue, and are clearly better.} 
While the naive method yields more than 10 missing true matches in $58\%$ of time, our new approach improve this to just $8\%$. Moreover, the sample mean of missing matches also improves to 5.79, which is a factor or 2 lower than the greedy.  
The plot also shows that the ILP algorithm provides perfect matches more than $5\%$ of the time. 

One thing notable in the plot is that, for the ILP results, mistakes tend to occur in pairs and odd number errors appear in only $2\%$ of iterations. This is actually caused by sparsity and uncertainty when generating the universe and catalogs. An odd number of errors can only occur when we associate observations corresponding to at least 3 or more close-by objects. When the universe is sparse, it is more likely to have a pair of close objects than three or more objects close to each other, thus making an odd number of mismatches much more unlikely than an even number. 

\textcolor{black}{
We note that the above test as well as the following two assume perfect deblending, the ideal (yet impossible) scenario that corresponds to the worst case of source crowding. If the sources are blended on the images to the limit that the source extraction procedures cannot distinguish them, then both the greedy and the ILP methods improve in performance. 
This is studied in Section~\ref{sec:blending} in detail.
}

\subsection{Variable Exposure Depths}
Survey telescopes typically obtain observations with the same exposure times but the seeing (point-spread function) changes from time to time, which yields a variable depth of the extracted source lists.
We model this scenario by setting upper bounds of the selection intervals to 0.9, 0.95 and 1, while keeping the lower bounds at 0. 
\color{black}
The Venn diagrams inside Figure~\ref{fig:testD} illustrate the different subsets due to the selection functions. 

Figure~\ref{fig:testD}  demonstrates the significant improvement of the ILP algorithm. In the left panel, green error bars illustrate that for estimating 2-way matches, out of 180 true 2-way matches,  the new algorithm achieves perfect 2-way matches $65\%$ of the time, while the naive approach, shown by cyan error bars, gets perfect matches less than $10\%$ of the time. The sample mean also improves from $2.6$ to $0.4$, resulting in a drastic change in their approximated Poisson distribution presented by green and cyan solid bars respectively. The right panel also shows better performance of ILP approach in terms of the reported 3-way matches, whose sample mean is 5.6 comparing with 10.5 from the naive method. Moreover,  out of 3240 true 3-way matches, our procedure has less than 10 missing true matches over $90\%$ of the time. In comparison, the naive method has this accuracy only $50\%$ of the time.

\begin{figure*}
\centering
\hspace*{-1cm}\includegraphics[width = 1.1\textwidth]{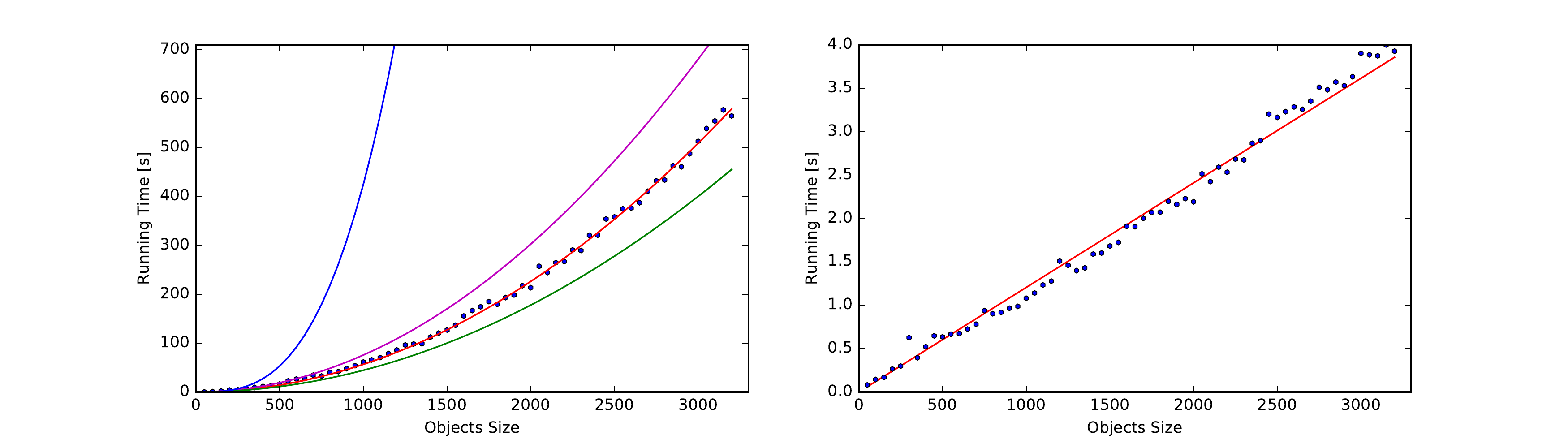}  
\caption{The runtime as a function of the size of the input catalog. \textit{Left}: The {blue} dots show the measured wall-clock time of the entire procedure: the red, magenta and green curves illustrate quadratic scaling with different coefficients, 
while blue curve shows a cubic relation for comparison.
\textit{Right}: The dots show the optimization runtime without preprocessing, which is well fit by a linear function. 
}

\label{fig:cost}
\end{figure*}
\begin{figure*}
\centering
\hspace*{-1cm}\includegraphics[width = 1.1\textwidth]{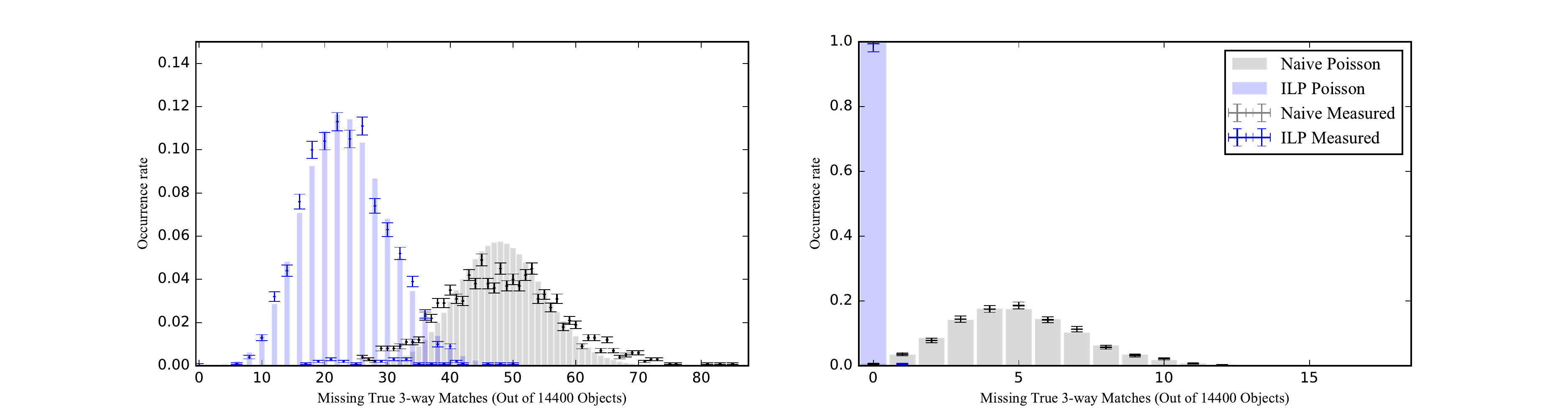} 
\caption{
Performance of the greedy (\textit{gray points with error bars}) and the optimal (\textit{blue}) algorithms in the limiting cases with surface density 400/sq.arcmin simulated in a field of view of 36 sq.arcminutes; the solid bars show the best fit Poisson distributions for reference.
The \textit{left} panel illustrates the (impossible) scenario where all sources are perfectly deblended, such that the catalogs are completely crowded, which makes the association very hard. The \textit{right} panel shows the scenario where we accept the fact that some blended sources will remain blended below a separation threshold, and hence we accept those as good matches, see text.
The truth is expected to be between these two extremes.
}
%

\label{fig:blend}
\end{figure*}

\subsection{The General Case of Different Populations}

Finally, we test the method's accuracy in a more complex scenario that simulates completely different instruments and their observations with the different selection functions. We model this by different overlapping intervals producing 1080 2-way matches and 2050 3-way matches. 
Figure~\ref{fig:testS} presents these results along with the schematics of the selection functions using the Venn diagrams. 
The 2-way matches on the left show that out of 1080 true 2-way matches, the ILP approach produces perfect  matches around $10\%$ of the time. The plot also shows the approximating Poisson distribution with sample mean $\lambda=2.7$. In contrast, the nearest neighbor method includes no perfect matches and fits a flatter Poisson distribution with $\lambda=16$. The right plot for 3-way matches demonstrates that when applying the new procedure, out of 2520 true 3-way matches, around $4.5\%$ of outputs are perfect matches and only $3.5\%$ of results have more than 10 missing matches. The naive approach reports no perfect matches and produces more than 10 missing matches $30\%$ of times.

\section{Discussion}
\label{sec:disc}

The algorithm presented here is the first attempt at solving the general catalog-level association problem for multiple catalogs with provable guarantees. An objective function based on the marginal likelihood is constructed, which is then optimized to global optimality by modeling it as an integer linear program. 
In the previous sections we see a significant improvement due to the global optimization approach over the naive greedy method.
Next we will discuss the feasibility of the approach by studying its performance and computational speed. 
Furthermore, we look at the limit of imperfect deblending.

\subsection{Runtime Analysis}

For practical use, the efficiency is considered as key to applicability. \citet{BB16-hungarian} show that, for 2-way catalog matching, we can sparsify the problem by eliminating variables corresponding to highly unlikely associations, which in turn cuts down significantly the runtime. The same holds for our new algorithm as well. In a direct comparison of the wall-clock times, we find that our ILP approach method is over a factor of 50 faster than the Hungarian algorithm, which is most encouraging for large-scale applications.
Of course, these differences could very well depend on the implementations -- a more efficient implementation of the Hungarian algorithm should be more comparable to the ILP solver. Moreover, with increasing number of objects, we expect the Hungarian algorithm to overtake the ILP approach in the 2-way case. 

For the 3-catalog case discussed in this paper, the new algorithm has even more impressive performance. As seen in the left panel of Figure~\ref{fig:cost}, the running time curve of the whole procedure, including pruning and optimization, scales quadratically with number of observations. 
This is primarily due to the inefficiency of our pruning procedure, which was not the focus of our study, as many approaches exist to speed this up, e.g., \citet{htm, healpix, zones, matthias2017gpu}.
In the right panel, we plot the runtime for the optimization only, which shows a clear linear scaling with number of observations. 

\subsection{Imperfect Deblending} \label{sec:blending}

The previous tests assumed perfect deblending, which is both the ideal scenario for source detection and the worst case for crowding in the resulting catalogs.
Next we look at how the results change in case of blended sources as an illustration of the expected changes.

We adopt an approximate heuristic formula for our simulations based on the characteristics of extracted sources in the Sloan Digital Sky Survey using a nominal astronomical uncertainty of \mbox{$\sigma\!\approx\!0.1"$} and a point-spread function (PSF) of \mbox{FWHM $\approx 1.4"$}. Keeping to this ratio, we scale to our previously assumed \mbox{$\sigma\!=\!0.04"$}, which yields a PSF \mbox{FWHM $\approx 0.5"$}. 
This corresponds to about 2.5 pixels in LSST, our motivating example mentioned earlier.
If we assume that sources closer than that separation on the sky cannot be deblended by the photometric software stack, we can simulate this effect by modifying the mock universe to include only one of the two colliding objects. {\color{black} Accordingly, the source catalogs now again have at most one source per astrophysical object (e.g., star).} This eliminates the ambiguity of the original blending issue when multiple stars could have a single blended source or photometric measurement. Of course, this is overly optimistic but our goal here is to study that limit, in addition to the previous (pessimistic) situation for crossmatching, the case of perfect deblending. The true expected performance is clearly between these two idealized scenarios.

To draw statistically significant conclusions, we created simulations in a larger area of 36 sq.~arcmin, with the same object density as in section~\ref{sec:accuracy}, and re-run the tests using the selection functions described in section~\ref{identical}. 
%
Figure~\ref{fig:blend} illustrates the difference between a greedy algorithm and our optimal solutions in two extreme scenarios: in the case of perfect deblending and the case of blended sources. We elaborate below.

Perfect deblending results in sources very close to each other on the sky, and hence crowding is the most challenging for correctly identifying the associations. The left panel of Figure~\ref{fig:blend} shows the performance in this scenario for the two approaches: the globally optimal solver producing more than two times fewer incorrect matches. The measurements are shown with blue and gray errorbars for the ILP and greedy algorithms, respectively. In addition, we also plot for reference the best-fit Poisson distribution. 
In particular, the greedy approach produces 48 wrong associations out of 14,400 objects on average, but it produces fewer than 40 bad matches only $15\%$ of time. 
In comparison, the ILP procedure achieves this same accuracy over $98\%$ of time and has a sample mean of 23 for missing true associations.

We model the situation with blending as follows: objects behind each other that are separated by an angle smaller than the PSF FWHM are not separated based on astrometry. We essentially eliminate the background object(s) and every catalog has at most one source corresponding to the blended objects. 
This obviously makes crowding less severe. In this limit both cross-identification approaches perform better as demonstrated in the right panel of Figure~\ref{fig:blend}: the greedy algorithm produces 10 times better results than previously but only manages to produce perfect associations $0.6\%$ of the time, which is in sharp contrast with our ILP method that succeeds $99.5\%$ of the time. 

We expect the truth to lie between these two extremes but the fact that our novel optimal solver provides significantly better results in both cases suggests superior performance for real catalogs.

\color{black}

\section{Concluding Remarks and Future Work}

The approach based on the Bayesian probabilistic modeling and our proposed solution using Integer Linear Programming is potentially of broader applicability than what was discussed in the previous sections. 
First of all, the new algorithm can deal with $N$ catalogs simultaneously, as discussed in Section~\ref{sec:ILP}. Here we demonstrated the power of the technique on three catalogs.

Second of all, the ILP solver does not have to run on the entire cartesian product of all catalogs. Instead it is meant to use a list of possible candidates found by existing high-performance tools (e.g., those based on the Hierarchical Triangular Mesh, HEALPix, or the Zones Algorithm). The ILP solver is to only be invoked on the sources in crowded or ambiguous regions where multiple competing matching scenarios are present.

Finally, our proposed method is applicable not only to static point sources (illustrated in this paper) but also to any other scenarios where one can model the morphology of objects and even their changes or proper motion (see citations in Section~1). 
This is achieved by starting with the marginal likelihoods derived from the appropriate geometrical and physical models. 
Similarly, more advanced astrometric and population models will work, e.g., with elliptic errors or realistic clustering \citep{mallinar-ascom201783}, as long as the marginal likelihoods $\{\mlike_o\}$ can be calculated. 
In other words, the new approach is a natural add-on package for all existing matching tools.  
These properties above make it an ideal extension to existing pipelines and a must-have tool for modern surveys.


Our future work includes extensions to more complex settings, such as the radio morphology \citep{dongwei2015radio}, where even 2-way catalog matches might comprise of up to 4 sources 
(Fan et al. 2018, in preparation).
%
On a conceptual level, further improvement is expected from going beyond the current maximum likelihood method. One can formulate simple priors for the different partition schemes by counting the number of associations with different number of constituents. Instead of treating these as priors, one can also consider these population probabilities in a hierarchical model, and infer them simultaneously.

\color{black}

\acknowledgments

TB acknowledges partial support from NSF Grant AST-1412566 and NASA via the awards NNG16PJ23C and STScI-49721 under NAS5-26555. AB acknowledges support from NSF Grant CMMI-1452820.

\bibliographystyle{plainnat} 

\setcitestyle{authoryear,open={((},close={))}}
 
\bibliography{xmatch2b.bib}

\end{document}